\documentclass[onecolumn,usenatbib]{mn2e}
\usepackage{epsf}

\newcommand{\sD}{{\cal D}}
\newcommand{\sG}{{\cal G}}
\newcommand{\sL}{{\cal L}}

\title[Nonlinear bending waves in Keplerian accretion discs]{Nonlinear
  bending waves in Keplerian accretion discs}

\author[G. I. Ogilvie]
  {G. I. Ogilvie$^{1,2}$\\
  $^1$Department of Applied Mathematics and Theoretical Physics,
  University of Cambridge, Centre for Mathematical Sciences,\\
  Wilberforce Road, Cambridge CB3 0WA\\
  $^2$Institute of Astronomy, University of Cambridge, Madingley Road,
  Cambridge CB3 0HA}

\begin{document}

\maketitle

\label{firstpage}

\begin{abstract}
  The nonlinear dynamics of a warped accretion disc is investigated in
  the important case of a thin Keplerian disc with negligible
  viscosity and self-gravity.  A one-dimensional evolutionary equation
  is formally derived that describes the primary nonlinear and
  dispersive effects on propagating bending waves other than
  parametric instabilities.  It has the form of a derivative nonlinear
  Schr\"odinger equation with coefficients that are obtained
  explicitly for a particular model of a disc.  The properties of this
  equation are analysed in some detail and illustrative numerical
  solutions are presented.  The nonlinear and dispersive effects both
  depend on the compressibility of the gas through its adiabatic index
  $\Gamma$.  In the physically realistic case $\Gamma<3$, nonlinearity
  does not lead to the steepening of bending waves but instead
  enhances their linear dispersion.  In the opposite case $\Gamma>3$,
  nonlinearity leads to wave steepening and solitary waves are
  supported.  The effects of a small effective viscosity, which may
  suppress parametric instabilities, are also considered.  This
  analysis may provide a useful point of comparison between theory and
  numerical simulations of warped accretion discs.
\end{abstract}


\section{Introduction}

The existence of warped accretion discs, in which the orbital plane of
the gas varies slowly with radius and possibly with time, is suggested
by both observational evidence and theoretical reasoning.  Of the
observational results, perhaps the best examples are certain X-ray
binary stars, including Her X-1, in which the X-ray source appears to
be periodically occulted by a precessing warped disc
(e.g.~\citealt{GB76}; \citealt{CCCL03}), and a few active galactic
nuclei, including NGC~4258, in which the warped shape of the disc is
revealed by maser emission (e.g.~\citealt{M95}; \citealt{G03}).
Theoretical arguments indicate that a fluid disc will be warped by an
external torque if it is misaligned with the equatorial plane of a
spinning black hole or magnetized star at its centre (\citealt{BP75};
\citealt{LS80}; \citealt{L99}), or with the orbital plane of a binary
companion, planet or other satellite (\citealt{PT95}).  Even in an
initially aligned system, the disc may develop a warp through a linear
instability of the coplanar state, depending on tidal, radiation or
magnetic forces (\citealt{L92}; \citealt{P96}; \citealt{L99}).
Interpreting the behaviour of warped discs is an important and
challenging problem in astrophysical fluid dynamics.

The basic aim of theoretical approaches in this subject has been to
understand how the shape of the disc evolves under the action of
internal stresses and external torques, and to derive practical
one-dimensional equations that describe this evolution.  When the warp
is sufficiently small the governing equations are linear and can be
obtained from a perturbation analysis of a flat disc.  \citet{PP83}
and \citet{PL95} derived linearized equations for warped discs,
superseding the pioneering but flawed analyses of \citet{BP75} and
\citet{P78}, and demonstrated that there are two basic dynamical
regimes for warped Keplerian discs in linear theory.  Let the disc
have angular velocity $\Omega(r)$ and vertical scale-height $H(r)$,
and let $\alpha$ denote the dimensionless effective viscosity
parameter of \citet{SS73}.  To a first approximation, if $\alpha\ga
H/r$ the warp satisfies a diffusion-type equation with an effective
diffusion coefficient $H^2\Omega/(2\alpha)$, larger by a factor of
$1/(2\alpha^2)$ than the effective kinematic viscosity of the disc.
On the other hand, if $\alpha\la H/r$ the warp satisfies a wave-type
equation and propagates with speed $H\Omega/2$.  The rapid diffusive
and non-dispersive wavelike behaviours are unique to Keplerian discs
as they result from a resonance between the orbital and epicyclic
frequencies that couples the vertical motion associated with the warp
to shearing horizontal epicyclic motions.

A possible limitation of these approaches is that the Eulerian
perturbation analysis on which the linearized equations are based
requires that the vertical displacement be small compared to $H$, and
is therefore formally invalid for any observable warp.  It is
appreciated that the linear theories may be valid for larger
amplitudes but this can only be demonstrated using a Lagrangian or
semi-Lagrangian method.  More recently it has also become possible to
test these theories and to study aspects of warped discs using
numerical simulations (e.g. \citealt{LNPT96}; \citealt{NP99}).  Such
simulations have focused mainly on the wavelike regime $\alpha\la H/r$
and have found reasonable agreement with \citet{PL95} for
small-amplitude warps.  There are significant discrepancies, however,
which suggest that nonlinear and dispersive effects can be important.

A particular concern has been that the shearing epicyclic motions
associated with the warp in either regime are predicted to be very
fast, comparable to or larger than the sound speed, for observable
warps.  Shocks may form if the warp has a short enough wavelength that
these motions collide \citep{NP99}, but the horizontal motions may
also be parametrically unstable and decay into smaller-scale inertial
waves (\citealt{PT95}; \citealt{GGO00}).  Numerical simulations
indicate that the shearing motions are also damped by
magnetohydrodynamic turbulence in a quasi-viscous manner
\citep{TOBPNS00}.  Further studies of all these effects are needed to
assess their importance.

In earlier work (\citealt{O99}; \citealt{O00}) I derived a fully
nonlinear theory of warped discs with effective viscosity, which
agrees with \citet{PP83} in the appropriate limit and also bears a
formal resemblance to the simplified nonlinear equations adopted by
\citet{P92}.  However, the theory of \citet{O99} does not describe the
regime of propagating bending waves in Keplerian discs with $\alpha\la
H/r$.  It was noted there that the nonlinear dynamics of the wavelike
regime is likely to be very different, and the purpose of the present
paper is to investigate this important case.

In Section~\ref{s:leading} of this paper I review the linear theory of
long-wavelength bending waves in Keplerian discs and introduce the
detailed derivation that follows in Section~\ref{s:derivation}.  The
nonlinear evolutionary equation is analysed and solved numerically in
Section~\ref{s:analysis} and conclusions are presented in
Section~\ref{s:conclusions}.

\section{Leading-order dynamics}

\label{s:leading}

The linearized equations for long-wavelength bending waves in a thin,
inviscid, Keplerian disc are (e.g. \citealt{LO00})
\begin{equation}
  \Sigma r^2\Omega{{\partial W}\over{\partial t}}=
  {{1}\over{r}}{{\partial\sG}\over{\partial r}},
\end{equation}
\begin{equation}
  {{\partial\sG}\over{\partial t}}=
  {{1}\over{4}}\Sigma H^2r^3\Omega^3{{\partial W}\over{\partial r}},
\end{equation}
where $W(r,t)$ and $\sG(r,t)$ are complex variables describing the
warp and the internal torque.  Specifically $W=l_x+{\rm i}l_y$ encodes
the horizontal Cartesian components of the unit vector ${\bmath l}$
normal to the plane containing an annulus of the disc at radius $r$
and time $t$, while $\sG=G_x+{\rm i}G_y$ does the same for the
horizontal internal torque in the disc (divided by $2\pi$).  Also
$\Omega(r)$ is the angular velocity, $\Sigma(r)$ is the surface
density and $H(r)$ is an effective density scale-height, defined by
\begin{equation}
  \Omega=\left({{GM}\over{r^3}}\right)^{1/2},\qquad
  \Sigma=\int_{-\infty}^\infty\rho\,{\rm d}z,\qquad
  \Sigma H^2=\int_{-\infty}^\infty\rho z^2\,{\rm d}z,
\end{equation}
where $M$ is the central mass, $\rho$ is the density and $z$ is the
vertical coordinate in the unperturbed disc.  For a vertically
isothermal disc $H$ is the usual Gaussian scale-height.

When $\sG$ is eliminated we obtain
\begin{equation}
  {{\partial^2W}\over{\partial t^2}}={{GM}\over{4}}{{1}\over{\Sigma r^{3/2}}}
  {{\partial}\over{\partial r}}
  \left(\Sigma^2H^2{{1}\over{\Sigma r^{3/2}}}
  {{\partial W}\over{\partial r}}\right),
\end{equation}
which implies that linear bending waves propagate non-dispersively at
speed $H\Omega/2$ \citep{PL95}.  If the model of the disc is such that
the product $\Sigma H$ is independent of $r$, we obtain the classical
wave equation
\begin{equation}
  {{\partial^2W}\over{\partial t^2}}={{\partial^2W}\over{\partial x^2}}
\end{equation}
in the conveniently rescaled spatial coordinate
\begin{equation}
  x=\int\left({{GM}\over{4}}\Sigma^2H^2\right)^{-1/2}\Sigma r^{3/2}\,{\rm d}r
  =2\int{{{\rm d}r}\over{H\Omega}}.
\end{equation}
With plausible applications and numerical simulations in mind, we
consider the case $H/r=\epsilon={\rm constant}$; then $\Sigma\propto
r^{-1}$ and
\begin{equation}
  x={{4}\over{3\epsilon\Omega}}\propto r^{3/2}.
\end{equation}
An outwardly propagating bending wave in this theory has the simple form
\begin{equation}
  W=\epsilon f(x-t),\qquad
  A=-r\Omega f(x-t),
\label{wave}
\end{equation}
where $A=2\sG/(\Sigma H^2r\Omega)$ is related to the radial velocity
amplitude through
\begin{equation}
  u_r={\rm Re}\left({{Az}\over{r}}\,{\rm e}^{-{\rm i}\phi}\right).
\label{ur}
\end{equation}
The function $f$ is unconstrained except by the initial conditions,
and the wave propagates without change of form.

This long-wavelength theory of bending waves is subject to small
corrections depending on the quantity $kH$, where $k$ is the radial
wavenumber.  \citet{LP93} analysed the linear adiabatic wave modes in
a vertically isothermal accretion disc in the case $|kr|\ll 1$.  The
bending wave of interest corresponds to $m=1$ and $n=0$ in their
notation, and its dispersion relation in a Keplerian disc can be
expanded in the form
\begin{equation}
  \frac{\omega}{\Omega}=\pm\frac{1}{2}kH-\left(\frac{3\Gamma-2}{8\Gamma}\right)(kH)^2+O(kH)^3,
\label{lp_dispersion}
\end{equation}
where $\Gamma$ is the adiabatic index.  The first term on the
right-hand side corresponds to the approximation of non-dispersive
propagation as obtained in the long-wavelength bending-wave theory,
while the second term indicates the primary dispersive correction.
The dispersion depends on the compressibility of the fluid and is a
weak effect if the wavelength is long compared to $H$.

Nonlinearity will also modify the above theory.  If the nonlinearity
is weak and of the same order of magnitude as the linear dispersion,
we may expect a similar outwardly propagating wave solution to
equation (\ref{wave}) to exist, except that $f$ will no longer be an
arbitrary function determined only by the initial conditions, but will
evolve according to a nonlinear evolutionary equation.  The situation
is analogous to the classical problem of long water waves, where weak
nonlinearity and dispersion combine in the Korteweg--de Vries (KdV)
equation (e.g. \citealt{W74}).  In that case nonlinearity and
dispersion have competing tendencies and solitary waves (specifically,
solitons) are found.  Another well known example is the nonlinear
Schr\"odinger (NLS) equation, which arises in diverse fields including
nonlinear optics.  If the sign of the nonlinear term is such as to
steepen waves, the NLS equation also admits solitons.  The KdV and NLS
equations are recognised as generic nonlinear equations that arise in
many applications.

Since the bending wave propagates through an inhomogeneous medium, it
is not expected in general that its evolutionary equation will have
constant coefficients.  However, it is found below that for the
self-similar disc model in which $H\propto r$ and $\Sigma\propto
r^{-1}$, which has reasonable scalings and is well suited for
numerical simulations, the coefficients of the equation can indeed be
made constant by an appropriate choice of coordinates.

\section{Derivation of the evolutionary equation}

\label{s:derivation}

\subsection{Basic equations}

The equations governing the dynamics of an ideal, compressible fluid
are the equation of mass conservation,
\begin{equation}
  {\rm D}\rho=-\rho\nabla\cdot{\bmath u},
\end{equation}
the adiabatic condition,
\begin{equation}
  {\rm D}p=-\Gamma p\nabla\cdot{\bmath u},
\end{equation}
and the equation of motion,
\begin{equation}
  {\rm D}{\bmath u}=-\nabla\Phi-{{1}\over{\rho}}\nabla p,
\end{equation}
where ${\bmath u}$ is the velocity, ${\rm D}=\partial_t+{\bmath
  u}\cdot\nabla$ is the Lagrangian time-derivative, $p$ is the
pressure and $\Phi$ is the gravitational potential.  For a polytropic
gas $\Gamma$ is a constant.

Consider a non-self-gravitating, Keplerian disc around a point mass
$M$.  The time-dependent warping of the disc is best described in a
nonlinear regime using the warped spherical polar coordinate system
$(r,\theta,\phi)$ introduced by \citet{O99}.  In this scheme the
warped mid-plane of the disc is defined by the coordinate surface
$\theta=\pi/2$, and the warp consists of a tilt angle $\beta(r,t)$
together with a twist angle $\gamma(r,t)$.  The Cartesian components
of the unit tilt vector are given by ${\bmath
  l}=(\sin\beta\cos\gamma,\sin\beta\sin\gamma,\cos\beta)$.

When expressed in warped coordinates, the governing equations become
\begin{equation}
  {\rm D}\rho=-\rho\left[{{1}\over{r^2}}\partial_r(r^2v_r)+
  {{1}\over{r\sin\theta}}\partial_\theta(v_\theta\,\sin\theta)+
  {{1}\over{r\sin\theta}}\partial_\phi v_\phi\right],
  \label{drho}
\end{equation}
\begin{equation}
  {\rm D}p=-\Gamma p\left[{{1}\over{r^2}}\partial_r(r^2v_r)+
  {{1}\over{r\sin\theta}}\partial_\theta(v_\theta\,\sin\theta)+
  {{1}\over{r\sin\theta}}\partial_\phi v_\phi\right],
  \label{dp}
\end{equation}
\begin{equation}
  {\rm D}u_r-{{u_\theta^2}\over{r}}-{{u_\phi^2}\over{r}}=-{{GM}\over{r^2}}-
  {{1}\over{\rho}}\sD p,
  \label{dur}
\end{equation}
\begin{equation}
  {\rm D}u_\theta+{{u_ru_\theta}\over{r}}-
  {{u_\phi}\over{r\sin\theta}}\left[u_\phi\cos\theta+
  r({\rm D}\beta)\sin\phi-r({\rm D}\gamma)\sin\beta\cos\phi\right]=
  -{{1}\over{\rho r}}\partial_\theta p,
  \label{dut}
\end{equation}
\begin{equation}
  {\rm D}u_\phi+{{u_ru_\phi}\over{r}}+
  {{u_\theta}\over{r\sin\theta}}\left[u_\phi\cos\theta+
  r({\rm D}\beta)\sin\phi-r({\rm D}\gamma)\sin\beta\cos\phi\right]=
  -{{1}\over{\rho r\sin\theta}}\partial_\phi p,
  \label{dup}
\end{equation}
where $(v_r,v_\theta,v_\phi)$ are the components of velocity relative
to the moving coordinate system,
\begin{equation}
  {\rm D}=\partial_t+v_r\partial_r+{{v_\theta}\over{r}}\partial_\theta+
  {{v_\phi}\over{r\sin\theta}}\partial_\phi
\end{equation}
is the Lagrangian time-derivative,
\begin{equation}
  \sD=\partial_r-\left[(\partial_r\beta)\cos\phi+(\partial_r\gamma)\sin\beta\sin\phi\right]\partial_\theta-\left[-(\partial_r\beta)\cos\theta\sin\phi+(\partial_r\gamma)
  (\cos\beta\sin\theta+\sin\beta\cos\theta\cos\phi)\right]
  {{1}\over{\sin\theta}}\partial_\phi
\end{equation}
is a modified radial derivative, and $(u_r,u_\theta,u_\phi)$ are the
absolute velocity components
\begin{equation}
  u_r=v_r,
\end{equation}
\begin{equation}
  u_\theta=v_\theta+r({\rm D}\beta)\cos\phi+r({\rm D}\gamma)\sin\beta\sin\phi,
\end{equation}
\begin{equation}
  u_\phi=v_\phi-r({\rm D}\beta)\cos\theta\sin\phi+
  r({\rm D}\gamma)(\cos\beta\sin\theta+\sin\beta\cos\theta\cos\phi).
\end{equation}
These equations involve no further approximation and are valid for
arbitrary warps.  As in \citet{O99}, meaningful dynamical equations
for $\beta$ and $\gamma$ can be obtained only if the disc is thin and
is defined to lie close to the surface $\theta=\pi/2$.  This
constraint is implied by the asymptotic analysis that follows.

\subsection{Asymptotic expansions}

We utilize the small parameter $\delta\ll1$, such that the angular
semi-thickness of the disc is $H/r=\epsilon=\delta^2$.  The equations
are to be expanded in a region of the $(r,t)$ plane that corresponds
to a neighbourhood of the line $x=t$ followed by the nominal centre of
an outwardly propagating bending wave at leading order.  (The problem
of an inwardly propagating wave can be considered using the
time-reversal symmetry of the problem.)  We therefore introduce the
scaled variables $\zeta$, $\xi$ and $\tau$ to resolve the vertical
structure of the disc and the evolution of the wave.  These are
defined by
\begin{equation}
  \theta={{\pi}\over{2}}-\delta^2\zeta,\qquad
  x-t=\delta^{-1}\xi,\qquad
  t=\delta^{-2}\tau,
\end{equation}
and the equations are to be valid where $\zeta$, $\xi$ and $\tau$ are
$O(1)$.  Thus $\zeta$ is a scaled vertical coordinate relative to the
mid-plane of the thin disc, $\xi$ is a scaled horizontal coordinate
relative to the nominal centre $x=t$ of the travelling wave, and
$\tau$ is a scaled time coordinate that follows the solution over the
time-scale on which the radial location of the centre of the wave
changes by a factor of order unity.  The scaling of $\xi$ means that
at any one time the wave is described within a region of radial extent
comparable to the geometric mean of $r$ and $H$ (Figure~1).

\begin{figure*}
  \centerline{\epsfbox{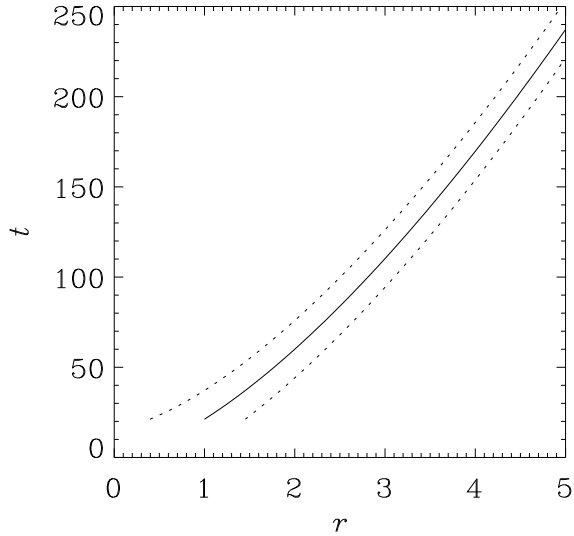}}
  \caption{Typical spatiotemporal domain of an outwardly propagating
    bending wave.  The solid line corresponds to the curve $x=t$ for a
    disc with $H/r=\delta^2=0.01$.  The dashed lines bound the region
    $|\xi|<10$ and indicate the region that might be occupied by a
    wave of modest radial extent.  Here $r$ is expressed in arbitrary
    units and $t$ in units of the orbital period at $r=1$.}
\end{figure*}

Partial derivatives transform according to
\begin{equation}
  \partial_t=-\delta\partial_\xi+\delta^2\partial_\tau,\qquad
  \partial_r=\delta^{-1}\frac{2}{r\Omega}\partial_\xi,\qquad
  \partial_\theta=-\delta^{-2}\partial_\zeta.
\end{equation}
The radius $r$ can be expressed in terms of $\xi$ and $\tau$
and expanded in the form
\begin{equation}
  r=2^{-4/3}3^{2/3}(GM)^{1/3}\tau^{2/3}
  \left[1+\delta\left({{2\xi}\over{3\tau}}\right)-
  \delta^2\left({{\xi^2}\over{9\tau^2}}\right)+\cdots\right]=r_0(\tau)+\delta r_1(\xi,\tau)+\delta^2r_2(\xi,\tau)+\cdots,
\end{equation}
and this allows any function of $r$ to be expanded similarly.  In
particular, the Keplerian angular velocity is
\begin{equation}
  \Omega=\left({{GM}\over{r^3}}\right)^{1/2}={{4}\over{3\epsilon x}}={{4}\over{3\tau}}\left[1-\delta\left({{\xi}\over{\tau}}\right)+
  \delta^2\left({{\xi^2}\over{\tau^2}}\right)+\cdots\right]=\Omega_0(\tau)+\delta\Omega_1(\xi,\tau)+\delta^2\Omega_2(\xi,\tau)+\cdots.
\end{equation}
We then propose the asymptotic expansions
\begin{equation}
  \beta=\delta^2\left[\beta_0(\xi,\tau)+\delta\beta_1(\xi,\tau)+\cdots\right],
  \label{beta_exp}
\end{equation}
\begin{equation}
  \gamma=\gamma_0(\xi,\tau)+\delta\gamma_1(\xi,\tau)+\cdots,
\end{equation}
\begin{equation}
  \rho=\delta^{2\sigma}\left[\rho_{\rm e}(r,\zeta)+
  \delta\rho_1(\phi,\zeta,\xi,\tau)+
  \delta^2\rho_2(\phi,\zeta,\xi,\tau)+\cdots\right],
\end{equation}
\begin{equation}
  p=\delta^{2\sigma+4}\left[p_{\rm e}(r,\zeta)+\delta p_1(\phi,\zeta,\xi,\tau)+
  \delta^2p_2(\phi,\zeta,\xi,\tau)+\cdots\right],
\end{equation}
\begin{equation}
  v_r=\delta\left[\delta v_{r1}(\phi,\zeta,\xi,\tau)+
  \delta^2v_{r2}(\phi,\zeta,\xi,\tau)+
  \delta^3v_{r3}(\phi,\zeta,\xi,\tau)+\cdots\right],
\end{equation}
\begin{equation}
  v_\theta=\delta^2\left[\delta v_{\theta1}(\phi,\zeta,\xi,\tau)+
  \delta^2v_{\theta2}(\phi,\zeta,\xi,\tau)+\cdots\right],
\end{equation}
\begin{equation}
  v_\phi=r(\Omega-{\rm D}\gamma)\sin\theta+
  \delta\left[\delta v_{\phi1}(\phi,\zeta,\xi,\tau)+
  \delta^2v_{\phi2}(\phi,\zeta,\xi,\tau)+
  \delta^3v_{\phi3}(\phi,\zeta,\xi,\tau)+\cdots\right].
  \label{v_phi_exp}
\end{equation}
Here $\rho_{\rm e}(r,\zeta)$ and $p_{\rm e}(r,\zeta)$ are the density
and pressure of the unperturbed disc in hydrostatic equilibrium,
which satisfy
\begin{equation}
  \partial_\zeta p_{\rm e}=-\rho_{\rm e}r^2\Omega^2\zeta.
  \label{hydrostatic_global}
\end{equation}
When expressed in terms of $\xi$ and $\tau$, these quantities have
expansions
\begin{equation}
  \rho_{\rm e}=\rho_0(\zeta,\tau)+\delta\rho_{{\rm e}1}(\zeta,\xi,\tau)+\cdots,
\end{equation}
\begin{equation}
  p_{\rm e}=p_0(\zeta,\xi,\tau)+\delta p_{{\rm e}1}(\zeta,\xi,\tau)+\cdots,
\end{equation}
and equation (\ref{hydrostatic_global}) is satisfied at every order in
$\delta$, in particular
\begin{equation}
  \partial_\zeta p_0=-\rho_0r_0^2\Omega_0^2\zeta.
\label{hydrostatic}
\end{equation}
The scaling of $\beta$ implies that the tilt angle of the disc is
$O(\delta^2)$ and therefore comparable to $H/r$, while the scaling of
$v_r$ implies that the radial velocity is comparable to the sound
speed.  It is found below that, with this choice of scalings, the
effect of nonlinearity in the solution is of comparable magnitude to
the effect of linear dispersion.  The indexing of the variables in
expansions (\ref{beta_exp})--(\ref{v_phi_exp}) is designed so as to
give a certain order to the equations that are deduced below.  The
parameter $\sigma$ is unspecified and drops out of the analysis.

\subsection{Basic structure of the disc}

When these expansions are substituted into equations
(\ref{drho})--(\ref{dup}), we obtain a number of equations to be
considered in turn.  Equation (\ref{dur}) at leading order [$O(1)$] is
satisfied, as the Keplerian rotation balances the gravitational force.
Equation (\ref{dut}) at leading order [$O(\delta^2)$] is satisfied by
virtue of the vertical hydrostatic equilibrium, equation
(\ref{hydrostatic}).

Equation (\ref{hydrostatic_global}) can be solved if the relation
between pressure and density is specified.  In order to make
analytical progress we consider an isothermal model for the vertical
structure, such that
\begin{equation}
  \rho_{\rm e}=\rho_{\rm m}\,{\rm e}^{-\zeta^2/2},
\end{equation}
\begin{equation}
  p_{\rm e}=p_{\rm m}\,{\rm e}^{-\zeta^2/2},
\end{equation}
where $\rho_{\rm m}(r)$ and $p_{\rm m}(r)$ are the (scaled) density
and pressure on the mid-plane.  In order to satisfy equation
(\ref{hydrostatic_global}), these quantities are related by
\begin{equation}
  p_{\rm m}=\rho_{\rm m}r^2\Omega^2.
\end{equation}

The surface density of the unperturbed disc is
\begin{equation}
  \Sigma=\delta^{2s+2}\left[\Sigma_0(\xi,\tau)+\cdots\right],
\end{equation}
with
\begin{equation}
  \Sigma_0=\int_{-\infty}^\infty\rho_0\,r_0\,{\rm d}\zeta=
  (2\pi)^{1/2}r_0\rho_0.
\end{equation}
In order that $\Sigma\propto r^{-1}$, as adopted in
Section~\ref{s:leading}, we require that $\rho_{\rm
  m}\propto r^{-2}$.

\subsection{Horizontal and vertical problems}

The equations derived at higher orders have a consistently repeating
formal structure that merits analysis.  The horizontal components of
the equation of motion, equations (\ref{dur}) and (\ref{dup}) at
$O(\delta^{n+1})$, $n\ge1$, involve the unknown quantities $v_{rn}$
and $v_{\phi n}$, and have the form
\begin{equation}
  \Omega_0\partial_\phi v_{rn}-2\Omega_0v_{\phi n}=F_{rn},
  \label{frn}
\end{equation}
\begin{equation}
  \Omega_0\partial_\phi v_{\phi n}+{{1}\over{2}}\Omega_0v_{rn}=F_{\phi n},
  \label{fphin}
\end{equation}
where the right-hand sides involve quantities known or partially known
from the lower orders.  These equations may be combined to give
\begin{equation}
  -2\Omega_0(\partial_\phi^2+1)v_{\phi n}=F_{{\rm h}n},
  \label{horizontal_n}
\end{equation}
where $F_{{\rm h}n}$ is the horizontal forcing combination
\begin{equation}
  F_{{\rm h}n}=F_{rn}-2\partial_\phi F_{\phi n}.
\end{equation}
The linear operator $\partial_\phi^2+1$, subject to periodic boundary
conditions in $\phi$, has eigenvalues $1-m^2$ and eigenfunctions ${\rm
  e}^{-{\rm i}m\phi}$, where $m$ is any integer.  The special case
$m=1$ is a null eigenfunction, or complementary function,
corresponding to an epicyclic motion or eccentric distortion of the
disc with an arbitrary dependence on $\zeta$.  The condition for
equation (\ref{horizontal_n}) to have a solution is
\begin{equation}
  \int_0^{2\pi}{\rm e}^{{\rm i}\phi}F_{{\rm h}n}\,{\rm d}\phi=0,
\end{equation}
i.e. that the forcing combination should contain no $m=1$ component.
(Since $F_{{\rm h}n}$ is real there is no difference in considering
$m=1$ or $m=-1$.)

The other three governing equations give rise to a vertical problem.
Equation (\ref{drho}) at $O(\delta^{2s+n})$, (\ref{dp}) at
$O(\delta^{2s+4+n})$ and (\ref{dut}) at $O(\delta^{2+n})$, $n\ge1$,
involve the unknown quantities $\rho_n$, $p_n$ and $v_{\theta n}$, and
have the form
\begin{equation}
  \Omega_0\partial_\phi\rho_n-{{v_{\theta n}}\over{r_0}}\partial_\zeta\rho_0-
  {{\rho_0}\over{r_0}}\partial_\zeta v_{\theta n}=F_{\rho n},
  \label{frhon}
\end{equation}
\begin{equation}
  \Omega_0\partial_\phi p_n-{{v_{\theta n}}\over{r_0}}\partial_\zeta p_0-
  {{\Gamma p_0}\over{r_0}}\partial_\zeta v_{\theta n}=F_{pn},
  \label{fpn}
\end{equation}
\begin{equation}
  \Omega_0\partial_\phi v_{\theta n}-{{1}\over{\rho_0r_0}}\partial_\zeta p_n+
  {{\rho_n}\over{\rho_0^2r_0}}\partial_\zeta p_0=F_{\theta n}.
\end{equation}
The quantities $\rho_n$ and $p_n$ may be eliminated, using the
hydrostatic condition (\ref{hydrostatic}), to obtain
\begin{equation}
  \sL v_{\theta n}=F_{{\rm v}n},
  \label{lvtn}
\end{equation}
where $\sL$ is a linear operator defined by
\begin{equation}
  \sL v_{\theta n}=-\partial_\zeta(\Gamma p_0\partial_\zeta v_{\theta n})+
  \rho_0r_0^2\Omega_0^2(\partial_\phi^2+1)v_{\theta n},
\end{equation}
and $F_{{\rm v}n}$ is the vertical forcing combination
\begin{equation}
  F_{{\rm v}n}=r_0^3\Omega_0^2\zeta F_{\rho n}+r_0\partial_\zeta F_{pn}+
  \rho_0r_0^2\Omega_0\partial_\phi F_{\theta n}.
\end{equation}

We define the eigenvalues $\lambda$ and eigenfunctions $w$ of $\sL$
as the solutions of the equation
\begin{equation}
  \sL w=\lambda\rho_0r_0^2\Omega_0^2w,
\end{equation}
subject to the conditions that $w$ have periodicity $2\pi$ in $\phi$
and be regular at the surfaces of the disc where the density and
pressure vanish.  With these boundary conditions $\sL$ is self-adjoint
with weight function $\rho_0$.  A vertically isothermal disc has no
definite surface and the vertical boundary condition is instead that
the wave energy flux tend to zero as $|\zeta|\to\infty$.  The
eigenfunctions and eigenvalues are then
\begin{equation}
  w_{\ell,m}={\rm e}^{-{\rm i}m\phi}\,{\rm He}_\ell(\zeta),\qquad
  \lambda_{\ell,m}=\ell\,\Gamma-m^2+1,
\end{equation}
where $\ell\ge0$ and $m$ are integers, and ${\rm He}$ denotes an
Hermite polynomial.\footnote{There are two definitions of Hermite
  polynomials.  Those used here are orthogonal with respect to the
  weight function $\exp(-\zeta^2/2)$.}  We will require only the first
four, ${\rm He}_0(\zeta)=1$, ${\rm He}_1(\zeta)=\zeta$, ${\rm
  He}_2(\zeta)=\zeta^2-1$ and ${\rm He}_3(\zeta)=\zeta^3-3\zeta$.

The special case $w_{0,1}={\rm e}^{-{\rm i}\phi}$, $\lambda_{0,1}=0$
is a null eigenfunction corresponding to a rigid tilt of the annulus
at radius $r$.  For general (irrational) $\Gamma$ this is the only
null eigenfunction.  The corresponding solvability condition on
equation (\ref{lvtn}) is
\begin{equation}
  \int_{-\infty}^\infty\int_0^{2\pi}{\rm e}^{{\rm i}\phi}F_{{\rm v}n}
  \,{\rm d}\phi\,{\rm d}\zeta=0.
\end{equation}

The forcing terms required for the development of the solution to the
desired order are listed in Appendix~A.

\subsection{Development of the solution}

Our objective is to determine the spatiotemporal devlopment of the
warp, and we must therefore obtain equations for
$\partial_\tau\beta_0$ and $\partial_\tau\gamma_0$ in terms of
$\beta_0$, $\partial_\xi\beta_0$, $\partial_\xi\gamma_0$, etc.  This
is done by solving the horizontal and vertical problems in turn up to
the required order and extracting the relevant solvability conditions.
Although straightforward in principle, this procedure involves arduous
algebraic manipulations and was carried out with the aid of
Mathematica.

The first horizontal problem is unforced: $F_{r1}=F_{\phi1}=0$.  The
general solution is an epicyclic motion (or eccentric distortion) with
an unknown dependence on position and time,
\begin{equation}
  v_{r1}=U_1(\zeta,\xi,\tau)\cos\phi+V_1(\zeta,\xi,\tau)\sin\phi.
\end{equation}
Based on our experience of the leading-order dynamics
(Section~\ref{s:leading}) we anticipate that in the absence of
viscosity the solution for an outwardly travelling bending wave will
be $U_1=-r_0\Omega_0\beta_0\zeta$ and $V_1=0$ (cf.
equations~\ref{wave} and~\ref{ur}).  This assumption simplifies the
calculation, and is verified subsequently.  Then we have
\begin{equation}
  v_{r1}=-r_0\Omega_0\beta_0\zeta\cos\phi,\qquad
  v_{\phi1}={{1}\over{2}}r_0\Omega_0\beta_0\zeta\sin\phi.
\label{vh1}
\end{equation}

The first vertical problem involves non-trivial forcing, and we find
\begin{equation}
  {{F_{{\rm v}1}}\over{\rho_0r_0^3\Omega_0^2}}=
  2(\Gamma-1)(\partial_\xi\beta_0\cos\phi+
  \beta_0\partial_\xi\gamma_0\sin\phi)(1-\zeta^2)-6\beta_0(\partial_\xi\beta_0\cos2\phi+
  \beta_0\partial_\xi\gamma_0\sin2\phi)\zeta,
\end{equation}
involving two different eigenfunctions, $w_{2,1}$ and $w_{1,2}$, of
the operator $\sL$.  The solvability condition is satisfied, and
equation (\ref{lvtn}) has the solution
\begin{equation}
  {{v_{\theta1}}\over{r_0}}=
  \left({{\Gamma-1}\over{\Gamma}}\right)
  (\partial_\xi\beta_0\cos\phi+\beta_0\partial_\xi\gamma_0\sin\phi)
  (1-\zeta^2)+\left({{6}\over{3-\Gamma}}\right)\beta_0
  (\partial_\xi\beta_0\cos2\phi+\beta_0\partial_\xi\gamma_0\sin2\phi)\zeta.
\end{equation}
(We do not add any multiple of the complementary function $w_{0,1}$,
as this would amount simply to redefining the tilt $\beta_1$.)  The
quantities $\rho_1$ and $p_1$ may then be obtained from equations
(\ref{frhon}) and (\ref{fpn}) after an integration with respect to
$\phi$, and we obtain
\begin{eqnarray}
  \lefteqn{\rho_1=\rho_1^{\rm ax}(\zeta,\xi,\tau)+
  {{\rho_0}\over{\Omega_0}}\left\{
  {{1}\over{\Gamma}}(\partial_\xi\beta_0\sin\phi-
  \beta_0\partial_\xi\gamma_0\cos\phi)
  \left[3-\Gamma+(\Gamma-1)\zeta^2\right]\zeta\right.}&\nonumber\\
  &&\left.\qquad+\left({{3}\over{3-\Gamma}}\right)\beta_0
  (\partial_\xi\beta_0\sin2\phi-\beta_0\partial_\xi\gamma_0\cos2\phi)
  (1-\zeta^2)\right\},
\end{eqnarray}
\begin{eqnarray}
  \lefteqn{p_1=p_1^{\rm ax}(\zeta,\xi,\tau)+
  {{p_0}\over{\Omega_0}}\left\{
  {{1}\over{\Gamma}}(\partial_\xi\beta_0\sin\phi-
  \beta_0\partial_\xi\gamma_0\cos\phi)
  \left[\Gamma+1+(\Gamma-1)\zeta^2\right]\zeta\right.}&\nonumber\\
  &&\left.\qquad+\left({{3}\over{3-\Gamma}}\right)\beta_0
  (\partial_\xi\beta_0\sin2\phi-\beta_0\partial_\xi\gamma_0\cos2\phi)
  (\Gamma-\zeta^2)\right\},
\end{eqnarray}
where the axisymmetric parts are required to satisfy
\begin{equation}
  \partial_\zeta p_1^{\rm ax}=-\rho_1^{\rm ax}r_0^2\Omega_0^2\zeta-
  \rho_0r_0^2\Omega_0\beta_0^2(\partial_\xi\gamma_0)\zeta.
\end{equation}
We may assume that $\rho_1^{\rm ax}$ vanishes, since it would
otherwise correspond to an arbitrary redefinition of the unperturbed
density of the disc.  Then
\begin{equation}
  p_1^{\rm ax}={{\beta_0^2(\partial_\xi\gamma_0)}\over{\Omega_0}}p_0.
\end{equation}

The second horizontal problem is also forced, and we find
\begin{equation}
  {{F_{{\rm h}2}}\over{r_0\Omega_0}}=
  -{{1}\over{2\Gamma}}\beta_0
  \left[\partial_\xi\beta_0\left(1+3\cos2\phi\right)+
  3\beta_0\partial_\xi\gamma_0\sin2\phi\right]
  \left[\Gamma-1+(\Gamma+1)\zeta^2\right]-\left({{12}\over{3-\Gamma}}\right)\beta_0^2
  (\partial_\xi\beta_0\cos3\phi+\beta_0\partial_\xi\gamma_0\sin3\phi)\zeta.
\end{equation}
Again, the solvability condition is satisfied.  (This verifies that
our choice for $U_1$ and $V_1$ was correct; if we had not assumed
their form in advance, the solvability condition at this order would
have implied equation~\ref{vh1}.)  Equation (\ref{horizontal_n}) then
has the solution
\begin{eqnarray}
  v_{\phi2}&=&
  {{1}\over{4\Gamma}}r_0\beta_0
  \left[\partial_\xi\beta_0\left(1-\cos2\phi\right)-
  \beta_0\partial_\xi\gamma_0\sin2\phi\right]
  \left[\Gamma-1+(\Gamma+1)\zeta^2\right]-{{3}\over{4(3-\Gamma)}}r_0\beta_0^2
  (\partial_\xi\beta_0\cos3\phi+\beta_0\partial_\xi\gamma_0\sin3\phi)\zeta
  \nonumber\\
  &&\qquad-{{1}\over{2}}U_2(\zeta,\xi,\tau)\sin\phi+
  {{1}\over{2}}V_2(\zeta,\xi,\tau)\cos\phi.
\end{eqnarray}
The last two terms are the complementary functions, presently of
unknown amplitude.  The quantity $v_{r2}$ may then be obtained from
equation (\ref{fphin}), and we find
\begin{eqnarray}
  v_{r2}&=&
  -{{1}\over{2\Gamma}}r_0\beta_0
  \left[\partial_\xi\beta_0\sin2\phi-
  \beta_0\partial_\xi\gamma_0(1+\cos2\phi)\right]
  \left[\Gamma-1+(\Gamma+1)\zeta^2\right]-
  {{3}\over{2(3-\Gamma)}}r_0\beta_0^2
  (\partial_\xi\beta_0\sin3\phi-\beta_0\partial_\xi\gamma_0\cos3\phi)\zeta
\nonumber\\
  &&\qquad-2r_0\beta_0^2(\partial_\xi\gamma_0)\zeta^2+\left[1-\left({{3}\over{3-\Gamma}}\right)\beta_0^2\right]r_0
  (\partial_\xi\beta_0\sin\phi-\beta_0\partial_\xi\gamma_0\cos\phi)\zeta+
  U_2\cos\phi+V_2\sin\phi.
\end{eqnarray}

The forcing for the second vertical problem is very complicated, but
fortunately it is not necessary to solve it in detail.  The
solvability condition alone provides an equation for the desired
quantities $\partial_\tau\beta_0$ and $\partial_\tau\gamma_0$.  This
involves much algebra, not written out here, and simplifies to
\begin{eqnarray}
  \lefteqn{\partial_\tau\beta_0+{\rm i}\beta_0\partial_\tau\gamma_0-
  \partial_\xi\beta_1-{\rm i}\beta_1\partial_\xi\gamma_0-
  {{1}\over{\Sigma_0r_0\Omega_0\,{\rm e}^{{\rm i}\gamma_0}}}
  \partial_\xi\int\rho_0(U_2+{\rm i}V_2)\,{\rm e}^{{\rm i}\gamma_0}\,
  r_0\zeta\,{\rm d}\zeta=-{{1}\over{4}}\Omega_0\beta_0-
  {{r_1}\over{2r_0}}(\partial_\xi\beta_0+{\rm i}\beta_0\partial_\xi\gamma_0)}&\nonumber\\
  &&-{{\rm i}\over{2\Omega_0}}\left[\partial_{\xi\xi}\beta_0-
  \beta_0(\partial_\xi\gamma_0)^2+{\rm i}\beta_0\partial_{\xi\xi}\gamma_0+
  2{\rm i}(\partial_\xi\beta_0)\partial_\xi\gamma_0\right]+{{\rm i}\over{2\Omega_0}}{{3(\Gamma-1)}\over{3-\Gamma}}
  \left[\beta_0^2\partial_{\xi\xi}\beta_0+
  2\beta_0(\partial_\xi\beta_0)^2\right]\nonumber\\
  &&-{{1}\over{\Omega_0}}{{3(\Gamma+1)}\over{3-\Gamma}}
  \beta_0^2(\partial_\xi\beta_0)\partial_\xi\gamma_0-{{1}\over{2\Omega_0}}\left({{\Gamma+3}\over{3-\Gamma}}\right)
  \beta_0^3\left[\partial_{\xi\xi}\gamma_0+
  {\rm i}(\partial_\xi\gamma_0)^2\right].
  \label{scv2}
\end{eqnarray}
Although this equation contains information about the evolution of
$\beta_0$ and $\gamma_0$, it also involves the unknown quantities
$\beta_1$, $\gamma_1$, $U_2$ and $V_2$.  Further information must
therefore be sought.

Fortunately it is not necessary to solve the second vertical problem
to obtain $v_{\theta2}$, $\rho_2$ and $p_2$ in detail.  The
contribution of these quantities to the third horizontal problem,
which is the last problem we consider, is of the form
\begin{equation}
  F_{{\rm h}3}=\cdots+{{v_{\theta2}}\over{r_0}}\partial_\zeta v_{r1}-
  2\partial_\phi\left({{v_{\theta2}}\over{r_0}}\partial_\zeta v_{\phi1}\right).
\end{equation}
For the solvability condition we require only the $m=1$ component of
this quantity.  In principle both the $m=0$ and $m=2$ components of
$v_{\theta2}$ could contribute, but the $m=2$ component gives exactly
zero contribution.  Now the axisymmetric part of $F_{{\rm v}2}$ is
\begin{equation}
  F_{{\rm v}2}^{\rm ax}=-\rho_0r_0^3\Omega_0\beta_0
  \left[2(\partial_\xi\beta_0)\partial_\xi\gamma_0+
  \beta_0\partial_{\xi\xi}\gamma_0\right]
  \left[\left({{2\Gamma^2+\Gamma-2}\over{\Gamma}}\right)\zeta+
  {{(\Gamma-1)(3\Gamma+1)}\over{\Gamma}}(\zeta^3-3\zeta)\right],
\end{equation}
involving the eigenfunctions $w_{1,0}$ and $w_{3,0}$ of $\sL$, and so
the axisymmetric part of $v_{\theta2}$ is
\begin{equation}
  v_{\theta2}^{\rm ax}=-{{r_0}\over{\Omega_0}}\beta_0
  \left[2(\partial_\xi\beta_0)\partial_\xi\gamma_0+
  \beta_0\partial_{\xi\xi}\gamma_0\right]
  \left[{{(2\Gamma^2+\Gamma-2)}\over{\Gamma(\Gamma+1)}}\zeta+
  \left({{\Gamma-1}\over{\Gamma}}\right)(\zeta^3-3\zeta)\right].
\end{equation}
This information is sufficient to deduce the final solvability
condition in the form
\begin{equation}
  (\partial_\tau\beta_0+{\rm i}\beta_0\partial_\tau\gamma_0+
  \partial_\xi\beta_1+{\rm i}\beta_1\partial_\xi\gamma_0)\zeta+
  {{1}\over{r_0\Omega_0\,{\rm e}^{{\rm i}\gamma_0}}}
  \partial_\xi\left[(U_2+{\rm i}V_2)\,{\rm e}^{{\rm i}\gamma_0}\right]=\cdots,
\end{equation}
where we do not write the right-hand side in full.  By multiplying
this equation by $\rho_0r_0\zeta/\Sigma_0$ and integrating with
respect to $\zeta$ we obtain
\begin{eqnarray}
  \lefteqn{\partial_\tau\beta_0+{\rm i}\beta_0\partial_\tau\gamma_0+
  \partial_\xi\beta_1+{\rm i}\beta_1\partial_\xi\gamma_0+
  {{1}\over{\Sigma_0r_0\Omega_0\,{\rm e}^{{\rm i}\gamma_0}}}
  \partial_\xi\int\rho_0(U_2+{\rm i}V_2)\,{\rm e}^{{\rm i}\gamma_0}\,
  r_0\zeta\,{\rm d}\zeta={{1}\over{4}}\Omega_0\beta_0+
  {{r_1}\over{2r_0}}(\partial_\xi\beta_0+{\rm i}\beta_0\partial_\xi\gamma_0)}&\nonumber\\
  &&+{{\rm i}\over{2\Omega_0}}\left({{7\Gamma-4}\over{\Gamma}}\right)\left[\partial_{\xi\xi}\beta_0-
  \beta_0(\partial_\xi\gamma_0)^2+{\rm i}\beta_0\partial_{\xi\xi}\gamma_0+
  2{\rm i}(\partial_\xi\beta_0)\partial_\xi\gamma_0\right]+{{\rm i}\over{2\Omega_0}}\left({{3}\over{3-\Gamma}}\right)
  \left[\beta_0^2\partial_{\xi\xi}\beta_0+
  (2-\Gamma)\beta_0(\partial_\xi\beta_0)^2\right]\nonumber\\
  &&-{{1}\over{\Omega_0}}{{(15+6\Gamma-\Gamma^2)}\over{(\Gamma+1)(3-\Gamma)}}
  \beta_0^2(\partial_\xi\beta_0)\partial_\xi\gamma_0-
  {{1}\over{2\Omega_0}}{{(\Gamma+9)}\over{(\Gamma+1)(3-\Gamma)}}
  \beta_0^3\left[\partial_{\xi\xi}\gamma_0+
  (\Gamma+1){\rm i}(\partial_\xi\gamma_0)^2\right],
  \label{sch3}
\end{eqnarray}
which may be compared with equation (\ref{scv2}).  The unknown
quantities $\beta_1$, $\gamma_1$, $U_2$ and $V_2$ may all be
eliminated by taking the average of the two equations, yielding the
desired evolutionary equation for $\beta_0$ and $\gamma_0$.  This is
expressed most compactly in terms of the complex tilt variable
$W_0=\beta_0{\rm e}^{{\rm i}\gamma_0}$:
\begin{equation}
  \partial_\tau W_0=
  {{3(3\Gamma-2)}\over{8\Gamma}}{\rm i}\tau\partial_{\xi\xi}W_0+
  {{9\Gamma}\over{16(3-\Gamma)}}{\rm i}\tau\left[
  a|W_0|^2\partial_{\xi\xi}W_0+(1-a)W_0^2\partial_{\xi\xi}W_0^*+bW_0^*(\partial_\xi W_0)^2+
  (1-b)W_0|\partial_\xi W_0|^2\right],
  \label{wtau}
\end{equation}
with dimensionless parameters
\begin{equation}
  a={{2(\Gamma^2+2\Gamma+3)}\over{3\Gamma(\Gamma+1)}},\qquad
  b={{\Gamma+6}\over{3\Gamma}},
\end{equation}
\begin{equation}
  1-a={{(\Gamma-3)(\Gamma+2)}\over{3\Gamma(\Gamma+1)}},\qquad
  1-b={{2(\Gamma-3)}\over{3\Gamma}}.
\end{equation}
We recall that the true complex tilt variable is $W(r,t)=\delta^2W_0(\xi,\tau)+O(\delta^3)$.

\section{Analysis of the evolutionary equation}

\label{s:analysis}

\subsection{Rescaling}

Equation (\ref{wtau}) is a complex nonlinear equation describing the
spatiotemporal development of the warp.  The first, linear term on the
right-hand side corresponds to the linear dispersion evident in
equation (\ref{lp_dispersion}) derived from \citet{LP93}.\footnote{An
  additional factor of $3$ appears because of the relations between
  $\tau$ and $\Omega$ and between $x$ and $r$.}  The nonlinear terms
are cubic and also depend on the compressibility of the gas.  The
appearance of the factor $(3-\Gamma)$ in the denominator is
significant and can be traced to the involvement of the mode
$w_{1,2}$, which is coupled resonantly if $\Gamma=3$, even if the disc
is not vertically isothermal \citep{O99}.  In reality $\Gamma=5/3$ is
the case of greatest relevance.  In discussing the properties of
equation (\ref{wtau}), however, it is instructive to contrast the
cases of $\Gamma<3$ and $\Gamma>3$.  We assume throughout that
$\Gamma\ge1$.

There is an explicit dependence on $\tau$ in equation (\ref{wtau})
because the bending wave experiences a varying background as it
propagates outwards through the disc.  However, this dependence can be
conveniently eliminated by working with the rescaled time variable
\begin{equation}
  T={{3(3\Gamma-2)}\over{16\Gamma}}\tau^2.
\end{equation}
The reason that an equation with constant coefficients can be obtained
is related to the self-similarity of the disc model.  The nonlinear
terms can also be rescaled by defining
\begin{equation}
  W_0=\left[{{2(3\Gamma-2)(3-\Gamma)}\over{3\Gamma^2}}\right]^{1/2}\Psi.
\end{equation}
We then have
\begin{equation}
  -{\rm i}\partial_T\Psi=\partial_{\xi\xi}\Psi+s\left[a|\Psi|^2\partial_{\xi\xi}\Psi+(1-a)\Psi^2\partial_{\xi\xi}\Psi^*+b\Psi^*(\partial_\xi\Psi)^2+(1-b)\Psi|\partial_\xi\Psi|^2\right],
\label{dnls}
\end{equation}
where
\begin{equation}
  s={\rm sgn}(3-\Gamma)=\pm1.
\end{equation}
Equation (\ref{dnls}) is a derivative nonlinear Schr\"odinger equation
(DNLS).\footnote{The term DNLS refers to a wide class of equations.
  The present equation, in which the nonlinear terms contain second
  derivatives, is not the form of DNLS most commonly studied.}  The
irreducible dimensionless parameters of the equation, $a$ and $b$,
satisfy $a>1$ and $b>1$ for $1\le\Gamma<3$, while $2/3<a<1$ and
$1/3<b<1$ for $\Gamma>3$.

\subsection{Elementary properties}

Equation (\ref{dnls}) is invariant under translations of $T$ and $\xi$
and also under the `gauge transformation' $\Psi\mapsto\Psi\,{\rm
  e}^{{\rm i}\chi}$, which corresponds to a trivial rotation of the
coordinate system through an angle $\chi$ about the $z$-axis.  A
further symmetry is $\xi\mapsto-\xi$, which corresponds to a
`reflection' about the centre of the wave.  The coordinate $T$, and
the equation itself, are invariant under time reversal
($\tau\mapsto-\tau$).  The trivial solution $\Psi=0$ corresponds to a
flat disc.  However, the equation is not invariant under
$\Psi\mapsto\Psi+{\rm constant}$, which might appear to correspond to
an additional trivial rigid tilt of the disc.  The reason can be
traced to equation (\ref{wave}) where it is seen that the amplitude
$W_0$ (or $\Psi$) determines the non-trivial horizontal velocities as
well as the tilt, because the solution under consideration is a wave
travelling in one direction.

\subsection{Travelling-wave solutions}

It is readily verified that solutions exist in the form of uniform
travelling waves $\Psi\propto\exp({\rm i}\omega T-{\rm i}k\xi)$, where
$\omega$ and $k$ satisfy the nonlinear dispersion relation
\begin{equation}
  \omega=-\left(1+2sb|\Psi|^2\right)k^2.
\end{equation}
Such a wave is linearly stable to long-wavelength disturbances if
\begin{equation}
  (1+s|\Psi|^2)\left[1+s(2a-1)|\Psi|^2\right]\ge0,
\end{equation}
and to short-wavelength disturbances if
\begin{equation}
  4sb+(2a+b-1)^2|\Psi|^2\ge0.
\end{equation}
In the rescaled variables $\xi$ and $T$ the (negative) linear
dispersion coefficient is unity.  Since $2a-1>0$ and $b>0$ in
practice, the waves are stable if $\Gamma<3$ (i.e. $s=+1$) and the
nonlinear terms increase their dispersion.  Therefore it may be
expected that solitary waves are not supported, and this appears to be
confirmed by the following analysis.  However, if $\Gamma>3$ (i.e.
$s=-1$), the waves are unstable except when $|\Psi|^2>1/(2a-1)>1$, and
the nonlinearity counteracts the linear dispersion.

Following the usual analysis of NLS solitons (e.g.~\citealt{W74}), we
consider more general travelling-wave solutions of the form
\begin{equation}
  \Psi=F(X)\exp({\rm i}\omega T),\qquad
  X=\xi-cT,
\end{equation}
where $\omega$ is a frequency and $c$ is a wave speed to be
determined.  Note that $c$ is a slow speed in rescaled coordinates and
is relative to the basic wave speed of $H\Omega/2$.  Writing
$F=R\exp({\rm i}\Phi)$ with $R$ and $\Phi$ real, we then find
\begin{equation}
  (1+sR^2)R''+sRR'^2=(1+2sbR^2)R\Phi'^2-R(c\Phi'-\omega),
\label{rpp}
\end{equation}
\begin{equation}
  R\left[1+s(2a-1)R^2\right]\Phi''+2\left[1+s(2a+b-1)R^2\right]R'\Phi'=cR'.
\label{phipp}
\end{equation}
These equations are completely integrable.  Equation (\ref{phipp}) is
linear in $\Phi'$ and has the solution
\begin{equation}
  R^2\Phi'=\frac{sc}{2b}+h\left[1+s(2a-1)R^2\right]^{-q},
\end{equation}
where $h$ is an arbitrary constant and $q=b/(2a-1)$.  We have $7/9\le
q<1$ for $1\le\Gamma<3$, while $1<q<25/23$ for $\Gamma>3$.  Equation
(\ref{rpp}) may then be written in the form
\begin{equation}
  (1+sR^2)R''+sRR'^2=-\frac{{\rm d}V}{{\rm d}R},
\end{equation}
with
\begin{equation}
  V(R)=\frac{c^2}{8b^2R^2}+\frac{sch}{2bR^2}\left[1+s(2a-1)R^2\right]^{1-q}+\frac{h^2}{2R^2}\left[1+s(2a-1)R^2\right]^{1-2q}-\frac{1}{2}\omega R^2,
\end{equation}
and has the first integral
\begin{equation}
  \frac{1}{2}(1+sR^2)R'^2+V(R)=E={\rm constant}.
\label{energy}
\end{equation}
There is an obvious mechanical analogy with a particle (albeit of
variable inertia) moving in an effective potential.

A solitary wave would have $R\to0$ and $R'\to0$ as $X\to\pm\infty$.
The analogous particle would roll in a potential well between $R=0$
and $R=R_+>0$ such that $V(R_+)=V(0)$ and $V(R)<V(0)$ for $0<R<R_+$.
In this problem, as $R\to0$, the effective potential diverges as
$V\sim(c+2sbh)^2/8b^2R^2$, so this behaviour is possible only if
$h=-sc/2b$.  In this case
\begin{equation}
  V=\frac{c^2}{8b^2R^2}\left\{1-2\left[1+s(2a-1)R^2\right]^{1-q}+\left[1+s(2a-1)R^2\right]^{1-2q}\right\}-\frac{1}{2}\omega R^2,
\end{equation}
and $V$ tends to a constant as $R\to0$.  We exclude the trivial case
$c=0$.  It can then be shown that $V$ is a strictly concave function
of $R^2$ (i.e. $d^2V/d(R^2)^2<0$) when $\Gamma<3$ (i.e. $s=+1$ and
$q<1$).  Therefore it is impossible for a potential well to be formed,
and we conclude that solitary waves are not supported when $\Gamma<3$.

In contrast, when $\Gamma>3$ (i.e. $s=-1$ and $q>1$), $V$ is a
strictly convex function of $R^2$ that diverges to $+\infty$ as $R^2$
approaches $1/(2a-1)$, and a potential well is always formed.  The
squared amplitude $R_+^2$ can never exceed $1/(2a-1)$, so it appears
that solitary waves are supported for small amplitudes $R^2<1/(2a-1)$
and stable extended wavetrains for larger amplitudes $R^2>1/(2a-1)$.
An analytical approximation for the solitary waves in the case
$\Gamma>3$ is possible when they are of small amplitude $R_+\ll1$.  In
this limit
\begin{equation}
  V(R)-V(0)\approx\frac{bc^2}{8}R^2(R^2-R_+^2),
\end{equation}
with
\begin{equation}
  R_+^2=\frac{1}{b}\left(\frac{4\omega}{c^2}-1\right).
\end{equation}
The solution of equation (\ref{energy}) for $R^2\ll1$ is then
\begin{equation}
  R\approx R_+{\rm sech}(kX),\qquad
  k=\frac{1}{2}b^{1/2}cR_+,
\label{solitary}
\end{equation}
and furthermore $\Phi'\approx c/2$.  For a specified height ($R_+$)
and width ($1/k$), the frequency $\omega$ and wave speed $c$ are
determined by these relations.


\subsection{Numerical solutions}

Equation (\ref{dnls}) was solved numerically by applying a spatial
discretization on a uniform grid, based on second-order centred
differences, and solving the resulting coupled ordinary differential
equations in time using a fifth-order Runge--Kutta method with
adaptive stepsize.  Illustrative solutions for the case $\Gamma=5/3$
are shown in Fig.~2.  The initial profile is a real Gaussian of
relatively large amplitude.  The solution of the linear Schr\"odinger
equation, obtained when the nonlinear terms are neglected, can be
written in closed form and is shown in the right-hand panels; it
exhibits some linear dispersion.  When the nonlinear terms are
included, as shown in the left-hand panels, the wave broadens much
more rapidly and there is a greater emission of short waves.  (If
their wavelength is comparable to or shorter than $H$ it is likely
that they will behave differently from the predictions of equation
\ref{dnls}).  The computational domain is large enough that the
boundary conditions do not affect the solution significantly for the
ranges of $\xi$ and $T$ shown.  It was also confirmed numerically that
small-amplitude solitary waves in the case $\Gamma>3$ propagate
according to equation (\ref{solitary}).

\begin{figure*}
  \centerline{\epsfysize22cm\epsfbox{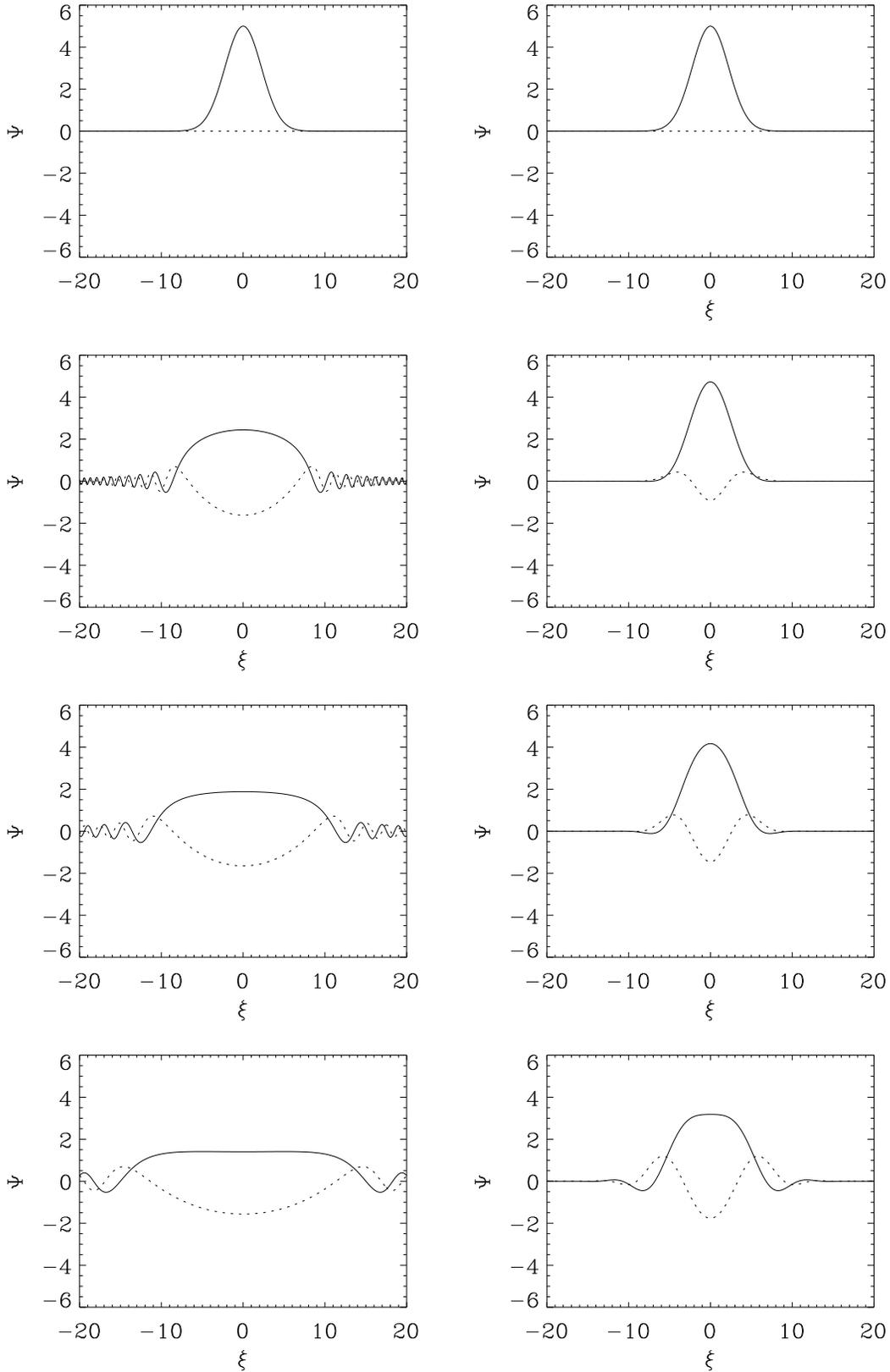}}
  \caption{Numerical solution of equation (\ref{dnls}) in the case
    $\Gamma=5/3$ starting from a real Gaussian initial condition.
    Left: solution including the nonlinear terms.  Right: solution
    excluding the nonlinear terms.  The real and imaginary parts of
    $\Psi$ are shown as solid and dotted lines, respectively, at times
    (from top to bottom) $T=0$, $1$, $2$ and $4$.  The computational
    domain is much larger than shown here.}
\end{figure*}

\subsection{Inclusion of a small effective viscosity}

It is straightforward to generalize the analysis to include a small
effective viscosity.  Suppose that the effective shear viscosity is
given by $\mu=\alpha p/\Omega$, where $\alpha$ is independent of $z$
and comparable to $H/r$.  Let $\alpha=\delta^2\alpha_0$ with
$\alpha_0=O(1)$.  Then the viscosity appears in the third-order
horizontal forcing terms $F_{r3}$ and $F_{\phi3}$, where its effect is
to replace $\partial_\tau$ by $(\partial_\tau+\alpha_0\Omega_0)$.
Indeed, any process that damps the shearing epicyclic motions at a
rate $\alpha\Omega$ would have an equivalent effect.  An effective
bulk viscosity of comparable magnitude would not affect the analysis.

The final equation for $W_0$ is modified to
\begin{equation}
  \partial_\tau W_0=\cdots-\frac{2\alpha_0}{3\tau}W_0,
\end{equation}
where the dots indicate the inviscid dynamical terms found previously.
The DNLS equation (\ref{dnls}) is modified to
\begin{equation}
  -{\rm i}\partial_T\Psi=\cdots+\frac{{\rm i}\alpha_0}{3T}\Psi.
\end{equation}
When $\Psi$ is small the effect of viscosity is to cause the solution
to decay by a factor $T^{-\alpha_0/3}$ relative to the inviscid
solution.  In general, however, the nonlinearity of the equation means
that it is not possible to find such an integrating factor.

\section{Conclusions}

\label{s:conclusions}

We have derived a one-dimensional equation describing the evolution of
weakly nonlinear and dispersive bending waves in a thin Keplerian
accretion disc with negligible viscosity and self-gravity.  The
nonlinear and dispersive effects both depend on the compressibility of
the gas through its adiabatic exponent $\Gamma$.  Perhaps
surprisingly, in the physically realistic case $\Gamma<3$,
nonlinearity tends to broaden the waves and enhance their linear
dispersion.  This is in contrast with the (hypothetical) case
$\Gamma>3$ in which nonlinearity counteracts linear dispersion and
solitary waves are supported.

Linear theories of warped accretion discs (\citealt{PP83};
\citealt{PL95}) previously made clear the distinction between the
diffusive ($\alpha\ga H/r$) and wavelike ($\alpha\la H/r$) regimes in
Keplerian discs.  Since they were based on an Eulerian perturbation
analysis they could not predict the amplitude at which they would
break down.  In this paper it is found that nonlinear effects cause a
significant modification of the bending wave (as the radial location
of the wave changes by a factor of order unity) when the dimensionless
amplitude of the warp $|{\rm d}W/{\rm d}\ln r|$ is $O(H/r)^{1/2}$ and
the horizontal velocities are comparable to the sound speed.

Unfortunately it is difficult to explain in simple terms how these
critical scalings arise or why the nonlinearity is steepening only
when $\Gamma>3$.  The nonlinearity arises through a mode-coupling
process and is complicated further by the resonant nature of the
epicyclic oscillations in a Keplerian disc, arising from the
coincidence of the orbital and epicyclic frequencies.

The analysis in this paper does not by any means present a complete
description of the nonlinear dynamics of the wavelike regime.  We have
focused on the evolution of a travelling bending wave and have not
considered the interaction between inward and outward waves, nor the
important case of steady warps maintained by an external torque.
Furthermore, \citet{NP99} noted that shocks can form if the wavelength
of the warp is short enough that the shearing epicyclic motions from
different parts of the wave collide.  This effect typically requires
$|{\rm d}W/{\rm d}\ln r|$ to be $O(1)$.  \citet{GGO00} also noted that
the warp could decay through a parametric instability.  The critical
amplitude for this effect is proportional to the effective viscosity
of the disc.

It would be interesting to compare the predictions of this analysis
with the results of numerical simulations of warped discs.  If
parametric instability sets in there should be a noticeable deviation
from equation~(\ref{dnls}).  The best way to make this comparison
would be to set up a vertically isothermal disc with $H\propto r$ and
$\Sigma\propto r^{-1}$, to introduce a warp and the accompanying
horizontal velocities (\ref{vh1}), and to follow the evolution of the
complex tilt $W(r,t)$.  The comparison with solutions of
equation~(\ref{dnls}) could then be made by transforming variables
from $W(r,t)$ to $\Psi(\xi,T)$, or vice versa.

\section*{Acknowledgements}

I acknowledge the support of the Aspen Center for Physics during the
research programme `Astrophysical Disks' (summer 2002), where this
work was mostly carried out, and of the Kavli Institute for
Theoretical Physics, Santa Barbara, during the research programme
`Physics of Astrophysical Outflows and Accretion Disks' (summer 2005),
where it was brought to completion.  This research was supported in
part by the National Science Foundation under grant number
PHY99-07949.

\appendix

\section{Forcing terms}

For reference, we list here the forcing terms required for the
development of the solution to the desired order.
\begin{equation}
  F_{\rho1}=-{{2\rho_0}\over{r_0\Omega_0}}[\partial_\xi v_{r1}-
  (\partial_\xi\gamma_0)\partial_\phi v_{r1}],
\end{equation}
\begin{eqnarray}
  \lefteqn{F_{\rho2}=-\partial_\tau\rho_0+
  \left(1-{{2v_{r1}}\over{r_0\Omega_0}}\right)\partial_\xi(\rho_{{\rm e}1}+\rho_1)-\left[\Omega_1+\left(1-{{2v_{r1}}\over{r_0\Omega_0}}\right)
  \partial_\xi\gamma_0\right]\partial_\phi\rho_1+
  {v_{\theta1}\over{r_0}}\partial_\zeta(\rho_{{\rm e}1}+\rho_1)}&\nonumber\\
  &&\qquad-
  {{r_1v_{\theta1}}\over{r_0^2}}\partial_\zeta\rho_0+(\rho_{{\rm e}1}+\rho_1)\left\{-{{2}\over{r_0\Omega_0}}[\partial_\xi v_{r1}-
  (\partial_\xi\gamma_0)\partial_\phi v_{r1}]+
  {{1}\over{r_0}}\partial_\zeta v_{\theta1}\right\}\nonumber\\
  &&\qquad+\rho_0\left\{-{{2}\over{r_0\Omega_0}}[\partial_\xi v_{r2}-
  (\partial_\xi\gamma_0)\partial_\phi v_{r2}-
  (\partial_\xi\gamma_1)\partial_\phi v_{r1}]\right.\nonumber\\
  &&\left.\qquad+
  \left({{2\Omega_1}\over{r_0\Omega_0^2}}+{{2r_1}\over{r_0^2\Omega_0}}\right)
  [\partial_\xi v_{r1}-(\partial_\xi\gamma_0)\partial_\phi v_{r1}]-
  {{2v_{r1}}\over{r_0}}-{{r_1}\over{r_0^2}}\partial_\zeta v_{\theta1}-
  {{1}\over{r_0}}\partial_\phi v_{\phi1}\right\},
\end{eqnarray}
\begin{equation}
  F_{p1}=\Gamma p_0\left\{-{{2}\over{r_0\Omega_0}}[\partial_\xi v_{r1}-
  (\partial_\xi\gamma_0)\partial_\phi v_{r1}]\right\},
\end{equation}
\begin{eqnarray}
  \lefteqn{F_{p2}=-\partial_\tau p_0+
  \left(1-{{2v_{r1}}\over{r_0\Omega_0}}\right)\partial_\xi(p_{{\rm e}1}+p_1)-\left[\Omega_1+\left(1-{{2v_{r1}}\over{r_0\Omega_0}}\right)
  \partial_\xi\gamma_0\right]\partial_\phi p_1+
  {v_{\theta1}\over{r_0}}\partial_\zeta(p_{{\rm e}1}+p_1)}&\nonumber\\
  &&\qquad-
  {{r_1v_{\theta1}}\over{r_0^2}}\partial_\zeta p_0+\Gamma(p_{{\rm e}1}+p_1)\left\{-{{2}\over{r_0\Omega_0}}[\partial_\xi v_{r1}-
  (\partial_\xi\gamma_0)\partial_\phi v_{r1}]+
  {{1}\over{r_0}}\partial_\zeta v_{\theta1}\right\}\nonumber\\
  &&\qquad+\Gamma p_0\left\{-{{2}\over{r_0\Omega_0}}[\partial_\xi v_{r2}-
  (\partial_\xi\gamma_0)\partial_\phi v_{r2}-
  (\partial_\xi\gamma_1)\partial_\phi v_{r1}]\right.\nonumber\\
  &&\left.\qquad+
  \left({{2\Omega_1}\over{r_0\Omega_0^2}}+{{2r_1}\over{r_0^2\Omega_0}}\right)
  [\partial_\xi v_{r1}-(\partial_\xi\gamma_0)\partial_\phi v_{r1}]-
  {{2v_{r1}}\over{r_0}}-{{r_1}\over{r_0^2}}\partial_\zeta v_{\theta1}-
  {{1}\over{r_0}}\partial_\phi v_{\phi1}\right\},
\end{eqnarray}
\begin{equation}
  F_{r1}=0,
\end{equation}
\begin{eqnarray}
  \lefteqn{F_{r2}=\left(1-{{2v_{r1}}\over{r_0\Omega_0}}\right)
  \partial_\xi v_{r1}+{{v_{\theta1}}\over{r_0}}\partial_\zeta v_{r1}-
  \left[\Omega_1+\left(1-{{2v_{r1}}\over{r_0\Omega_0}}\right)
  \partial_\xi\gamma_0\right]\partial_\phi v_{r1}+2\Omega_1v_{\phi1}}
  &\nonumber\\
  &&\qquad-{{2}\over{\rho_0r_0\Omega_0}}(\partial_\xi\beta_0\cos\phi+\beta_0\partial_\xi\gamma_0\sin\phi)\partial_\zeta p_0,
\end{eqnarray}
\begin{eqnarray}
  \lefteqn{F_{r3}=-\partial_\tau v_{r1}+
  \left(1-{{2v_{r1}}\over{r_0\Omega_0}}\right)
  \partial_\xi v_{r2}-\left({{2v_{r2}}\over{r_0\Omega_0}}-
  {{2v_{r1}r_1}\over{r_0^2\Omega_0}}-
  {{2v_{r1}\Omega_1}\over{r_0\Omega_0^2}}\right)\partial_\xi v_{r1}+
  {{v_{\theta1}}\over{r_0}}\partial_\zeta v_{r2}}&\nonumber\\
  &&\qquad+\left({{v_{\theta2}}\over{r_0}}-
  {{v_{\theta1}r_1}\over{r_0^2}}\right)\partial_\zeta v_{r1}-
  \left[\Omega_1+\left(1-{{2v_{r1}}\over{r_0\Omega_0}}\right)
  \partial_\xi\gamma_0\right]\partial_\phi v_{r2}\nonumber\\
  &&\qquad-\left[\Omega_2+{{v_{\phi1}}\over{r_0}}-\partial_\tau\gamma_0+
  \left(1-{{2v_{r1}}\over{r_0\Omega_0}}\right)\partial_\xi\gamma_1-
  \left({{2v_{r2}}\over{r_0\Omega_0}}-{{2v_{r1}r_1}\over{r_0^2\Omega_0}}-
  {{2v_{r1}\Omega_1}\over{r_0\Omega_0^2}}\right)\partial_\xi\gamma_0\right]
  \partial_\phi v_{r1}\nonumber\\
  &&\qquad+2\Omega_1v_{\phi2}+2\Omega_2v_{\phi1}+{{v_{\phi1}^2}\over{r_0}}-
  r_0\Omega_0^2\zeta^2\nonumber\\
  &&\qquad-{{2}\over{\rho_0r_0\Omega_0}}\left\{\partial_\xi(p_{{\rm e}1}+p_1)+
  (\partial_\xi\beta_0\cos\phi+\beta_0\partial_\xi\gamma_0\sin\phi)
  \partial_\zeta(p_{{\rm e}1}+p_1)\right.\nonumber\\
  &&\left.\qquad+
  [\partial_\xi\beta_1\cos\phi+(\beta_0\partial_\xi\gamma_1+
  \beta_1\partial_\xi\gamma_0)\sin\phi]\partial_\zeta p_0-
  (\partial_\xi\gamma_0)\partial_\phi p_1\right\}\nonumber\\
  &&\qquad+2\left({{\rho_{{\rm e}1}+\rho_1}\over{\rho_0^2r_0\Omega_0}}+
  {{r_1}\over{\rho_0r_0^2\Omega_0}}+{{\Omega_1}\over{\rho_0r_0\Omega_0^2}}\right)(\partial_\xi\beta_0\cos\phi+\beta_0\partial_\xi\gamma_0\sin\phi)
  \partial_\zeta p_0,
\end{eqnarray}
\begin{equation}
  F_{\theta1}=2(2v_{r1}-r_0\Omega_0)(\partial_\xi\beta_0\sin\phi-
  \beta_0\partial_\xi\gamma_0\cos\phi)-2(\partial_\phi v_{r1})
  (\partial_\xi\beta_0\cos\phi+\beta_0\partial_\xi\gamma_0\sin\phi),
\end{equation}
\begin{eqnarray}
  \lefteqn{F_{\theta2}=\left(1-{{2v_{r1}}\over{r_0\Omega_0}}\right)
  \partial_\xi v_{\theta1}+{{v_{\theta1}}\over{r_0}}\partial_\zeta v_{\theta1}-
  \left[\Omega_1+\left(1-{{2v_{r1}}\over{r_0\Omega_0}}\right)
  \partial_\xi\gamma_0\right]\partial_\phi v_{\theta1}}&\nonumber\\
  &&\qquad-\left({{r_1}\over{\rho_0r_0^2}}+{{\rho_{{\rm e}1}+\rho_1}\over{\rho_0^2r_0}}\right)
  \partial_\zeta p_1+{{\rho_1^2}\over{\rho_0^3r_0}}\partial_\zeta p_0+2\Omega_0v_{\phi1}\zeta\nonumber\\
  &&\qquad+2(2v_{r1}-r_0\Omega_0)[\partial_\xi\beta_1\sin\phi-
  (\beta_0\partial_\xi\gamma_1+\beta_1\partial_\xi\gamma_0)\cos\phi]\nonumber\\
  &&\qquad+2(2v_{r2}-r_0\Omega_1-r_1\Omega_0)(\partial_\xi\beta_0\sin\phi-
  \beta_0\partial_\xi\gamma_0\cos\phi)\nonumber\\
  &&\qquad-2(\partial_\phi v_{r1})
  [\partial_\xi\beta_1\cos\phi+(\beta_0\partial_\xi\gamma_1+
  \beta_1\partial_\xi\gamma_0)\sin\phi]+2r_0\Omega_0(\partial_\tau\beta_0\sin\phi-
  \beta_0\partial_\tau\gamma_0\cos\phi)\nonumber\\
  &&\qquad-{{2}\over{\Omega_0}}\left\{\Omega_0\partial_\phi v_{r2}-
  \left(1-{{2v_{r1}}\over{r_0\Omega_0}}\right)[\partial_\xi v_{r1}-
  (\partial_\xi\gamma_0)\partial_\phi v_{r1}]-
  {{v_{\theta1}}\over{r_0}}\partial_\zeta v_{r1}\right\}
  (\partial_\xi\beta_0\cos\phi+\beta_0\partial_\xi\gamma_0\sin\phi)\nonumber\\
  &&\qquad-
  r_0\left(1-{{2v_{r1}}\over{r_0\Omega_0}}\right)^2
  \{[\partial_{\xi\xi}\beta_0-\beta_0(\partial_\xi\gamma_0)^2]\cos\phi+
  [\beta_0\partial_{\xi\xi}\gamma_0+
  2(\partial_\xi\beta_0)\partial_\xi\gamma_0]\sin\phi\},
\end{eqnarray}
\begin{equation}
  F_{\phi1}=0,
\end{equation}
\begin{equation}
  F_{\phi2}=\left(1-{{2v_{r1}}\over{r_0\Omega_0}}\right)
  \partial_\xi v_{\phi1}+{{v_{\theta1}}\over{r_0}}\partial_\zeta v_{\phi1}-
  \left[\Omega_1+\left(1-{{2v_{r1}}\over{r_0\Omega_0}}\right)
  \partial_\xi\gamma_0\right]\partial_\phi v_{\phi1}+
  {{\Omega_1v_{r1}}\over{2}}+{{3\Omega_0r_1}\over{2r_0}}v_{r1},
\end{equation}
\begin{eqnarray}
  \lefteqn{F_{\phi3}=-\partial_\tau v_{\phi1}+
  \left(1-{{2v_{r1}}\over{r_0\Omega_0}}\right)
  \partial_\xi v_{\phi2}-\left({{2v_{r2}}\over{r_0\Omega_0}}-
  {{2v_{r1}r_1}\over{r_0^2\Omega_0}}-
  {{2v_{r1}\Omega_1}\over{r_0\Omega_0^2}}\right)\partial_\xi v_{\phi1}+
  {{v_{\theta1}}\over{r_0}}\partial_\zeta v_{\phi2}}&\nonumber\\
  &&\qquad+\left({{v_{\theta2}}\over{r_0}}-
  {{v_{\theta1}r_1}\over{r_0^2}}\right)\partial_\zeta v_{\phi1}-
  \left[\Omega_1+\left(1-{{2v_{r1}}\over{r_0\Omega_0}}\right)
  \partial_\xi\gamma_0\right]\partial_\phi v_{\phi2}\nonumber\\
  &&\qquad-\left[\Omega_2+{{v_{\phi1}}\over{r_0}}-\partial_\tau\gamma_0+
  \left(1-{{2v_{r1}}\over{r_0\Omega_0}}\right)\partial_\xi\gamma_1-
  \left({{2v_{r2}}\over{r_0\Omega_0}}-{{2v_{r1}r_1}\over{r_0^2\Omega_0}}-
  {{2v_{r1}\Omega_1}\over{r_0\Omega_0^2}}\right)\partial_\xi\gamma_0\right]
  \partial_\phi v_{\phi1}\nonumber\\
  &&\qquad+{{1}\over{2}}\Omega_1v_{r2}+2\Omega_2v_{r1}-
  {{v_{r1}v_{\phi1}}\over{r_0}}+
  {{3}\over{2}}\left[{{\Omega_0r_1}\over{r_0}}v_{r2}+
  \left({{2\Omega_0r_2}\over{r_0}}-{{\Omega_1^2}\over{\Omega_0}}-
  {{\Omega_0r_1^2}\over{r_0^2}}\right)v_{r1}\right].
\end{eqnarray}

\label{lastpage}

\end{document}